\documentclass[prl,twocolumn,nobalancelastpage]{revtex4}
\usepackage{amsmath}
\usepackage{graphicx}

\input{tcilatex}

\begin{document}

\title{Excitability in a nonlinear magnetoacoustic resonator}
\author{V. J. S\'{a}nchez--Morcillo, J. Redondo and J. Mart\'{\i}nez-Mora}
\affiliation{Departament de F\'{\i}sica Aplicada, Universitat Polit\`{e}cnica de Val\`{e}%
ncia, Crta. Natzaret-Oliva s/n, 46730 Grau de Gandia, Spain}

\begin{abstract}
We report a nonlinear acoustic system displaying excitability. The
considered system is a magnetostrictive material where acoustic waves are
parametrically generated. For a set of parameters, the system presents
homoclinic and heteroclinic dynamics, whose boundaries define a excitability
domain. The excitable behaviour is characterized by analyzing the response
of the system to different external stimuli. Single spiking and bursting
regimes have been identified.
\end{abstract}

\maketitle

There is an increasing interest in the study of smart materials \cite{smart}%
. The term smart refers to a class of materials that are highly responsive
and have the inherent capability to sense and react according to changes in
the environment. The most widely used classes of smart materials are
ferroic, and include piezoelectrics, electrostrictors, magnetostrictors, and
shape-memory alloys. In this letter we consider a smart system based on the
magnetostrictive effect. Magnetostriction is the material property that
causes a material to change its dimensions under the action of a magnetic
field or, inversely, to generate a magnetic field when deformed by an
external force. Magnetostrictive materials can thus be used for both sensing
and actuation.

One of the main goals of smart materials research is to design systems able
to mimic biological behaviour. For example, artificial muscles have been
recently developed based on electroactive polymers. Possibly the smartest
biological system is the neuron, responsible among others of the signal
processing in the brain. This processing capability relies to a large extend
on a property on the individual neurons called excitability \cite%
{Izhikevich00}. The most characteristic feature of excitability is a highly
nonlinear response (of all--or--nothing type) to an external stimulus: when
the amplitude of the stimulus is below a given threshold, there is a weak
reaction, while the response is strong, but independent of the strength of
the stimulation, above the threshold. The interest in excitable systems came
first from chemistry (reaction--diffusion systems) and biology (cardiac
tissue and neural modeling) \cite{Murray90}. More recently, excitability has
been identified in certain optical systems, as the CO$_{2}$ laser with
saturable absorber \cite{Plaza97} or the semiconductor laser with optical
feedback \cite{Giacomelli00}. In this letter we report on the first, to the
best of our knowledge, acoustic system displaying excitability. The system
also displays a rich complex behaviour including an scenario of
homoclinic/heteroclinic dynamics, which is on the basis of the excitable
character of the system.

The considered physical system consists in a magnetostrictive material
(e.g., a ferrite) in the form of a rod with transverse section $S$ and
length $L,$ with parallel and flat lateral boundaries, which is driven by an
external magnetic field $H_{ext}=H_{0}+H_{q}(t)$ parallel to the axis of the
rod. The oscillating magnetic field $H_{q}(t)=h\cos (2\omega t)$ induce a
temporal modulation of the sound velocity in the material at frequency $%
2\omega $, which is responsible for the parametric excitation of an acoustic
wave in the material at half of the driving frequency. A standing wave is
formed since the polished boundaries define a high-Q resonator for the sonic
beam. This effect has been considered in a number of works \cite%
{Brysev98,Brysev00,Preobrazhensky93}, where a radiofrequency magnetic field
was used to excite parametrically an ultrasonic wave. In these works, a
special attention was paid to the phenomenon of phase conjugation (PC) of an
incident acoustic beam, with important applications in e.g. acoustic
microscopy \cite{Brysev002}.

In the simplest scheme, the oscillating magnetic field is produced by a coil
with $n$ turns surrounding the material, The coil provides the inductance of
an electric RLC series circuit, driven by an ac source at frequency $2\omega 
$ and variable amplitude $\mathcal{E}$. Owing to magnetoelastic coupling,
elastic deformations in the magnetostrictive material result in an
additional magnetic field. Following \cite{Brysev98,Preobrazhensky93}, we
consider that the dominant contribution of the magnetoelastic interaction is
quadratic in the particle displacements $H_{int}(t)=-\alpha ~\left\langle u(%
\mathbf{r},t)^{2}\right\rangle $, where $\alpha =(2k)^{2}(\partial \ln
v/\partial H)$ is the coupling coefficient proportional to the modulation
depth of sound velocity \cite{Preobrazhensky93}, and the brackets indicate a
spatial average over the material volume. Taking into account the
magnetoelastic contribution, the effective magnetic excitation in the
material takes the form $H=H_{ext}(t)+H_{int}(t).$ The resulting magnetic
induction $B=\mu \left( H\right) H$ is in general a nonlinear relation,
which to the leading order can be written as $B=\mu H+\frac{1}{6}\mu
_{0}\chi ^{(3)}H^{3}$ \cite{Streltsov97}, where $\mu $ is the linear
permeability of the material and $\chi ^{(3)}$ the third order magnetic
susceptibility, which in turn depends on the frequency.

A nonlinear equation for the electric circuit can be obtained under the
previous assumptions, and neglecting nonresonant terms and those higher than
quadratic in $H_{q}$ and $H_{int}$. It reads%
\begin{align}
& \mathcal{L}\frac{d^{2}q}{dt^{2}}+R\frac{dq}{dt}+\frac{q}{C}=\mathcal{E}%
\cos (2\omega t)+  \notag \\
& \mu n\alpha \frac{d}{dt}\left\langle u^{2}\right\rangle +\mu _{0}\chi
^{(3)}n^{2}H_{0}\alpha \frac{d}{dt}\left( \frac{dq}{dt}\left\langle
u^{2}\right\rangle \right) ,  \label{charge}
\end{align}%
where $q$ is the charge in the capacitor, related with the current as $%
I=dq/dt$, $\mathcal{L}$ is the coil inductance and $H_{q}=nI.$ The last two
terms result from the nonlinearities related to magnetoelastic interaction
and magnetic nonlinearity.

The acoustic field obeys the wave equation with a source (coupling) term
proportional to the exciting magnetic field\emph{\ }\cite{Preobrazhensky93}.
In terms of the charge it reads 
\begin{equation}
\frac{1}{v^{2}}\frac{\partial ^{2}u}{\partial t^{2}}-\nabla ^{2}u=\alpha n%
\frac{dq}{dt}u.  \label{sound}
\end{equation}

We consider solutions of Eqs. (\ref{charge}) and (\ref{sound}) in the form
of quasi-harmonic waves, i.e., whose amplitudes are slowly varying in time.
In this case we can write 
\begin{subequations}
\begin{align}
q\left( t\right) & =\frac{1}{2}\left[ Q\left( t\right) \exp \left( 2i\omega
t\right) +c.c.\right] ,  \label{fieldQ} \\
u\left( \mathbf{r},t\right) & =\frac{1}{2}\left[ U\left( t\right) \exp
\left( i\omega t\right) +c.c.\right] g\left( \mathbf{r}_{\bot }\right) \sin
\left( kz\right) ,  \label{fieldU}
\end{align}%
where $u\left( \mathbf{r},t\right) $ is an eigenmode of the acoustical
resonator, $k=m\pi /L$ define the cavity resonances, and $2\omega =1/\sqrt{%
\mathcal{L}C}$ is the pumping frequency at the circuit resonance. In the
slowly-varying envelope approximation, the amplitudes obey the evolution
equations (see \cite{Sanchez05} for details) 
\end{subequations}
\begin{align}
\frac{dX}{d\tau }& =\mathcal{P}-X+Y^{2}+i\eta \left\vert Y\right\vert ^{2}X,
\label{eqs2} \\
\frac{dY}{d\tau }& =-\gamma \left( Y-XY^{\ast }\right) ,  \notag
\end{align}%
where $X$ and $Y$ are normalized values of the charge in the capacitor and
the amplitude of the ultrasonic field, respectively, which relate to the
physical variables through the transformations $X=\left( v^{2}n\alpha
/2\gamma _{U}\right) Q\ $and$\ Y=\frac{1}{4}\left( v\alpha n\right) (\mu L/%
\mathcal{L}\gamma _{U}\gamma _{Q})^{1/2}U.\ $The pumping term$\ \mathcal{P}%
=(v^{2}n\alpha /4\omega \gamma _{U}R)\mathcal{E}$ is proportional to the
voltage of the ac source. The parameter $\gamma =\gamma _{U}/\gamma _{Q}$
represents the ratio between losses, being $\gamma _{Q}=R/2\mathcal{L}\ $and 
$\gamma _{U}$ the electric and acoustic decay rates respectively. The last
parameter is introduced phenomenologically, and take into account the losses
due mainly to radiation from the boundaries. A dimensionless time $\tau
=\gamma _{Q}t$ has been also defined. Finally, $\eta =4\omega H_{0}\gamma
_{U}\chi ^{(3)}/\alpha v^{2}\mu _{r}$ remains as the single parameter of
nonlinearity, with $\mu _{r}=\mu /\mu _{0}$. Due to the number of parameters
involved in $\eta $, it can be varied over a wide range of values. Note that
Eqs. (\ref{eqs2}) also possess the $\emph{Z}_{2}$ (reflection) symmetry $%
(X,Y)\rightarrow (X,-Y)$.

The stationary solutions of Eqs. (\ref{eqs2}) and their stability have been
analysed in \cite{Sanchez05}. This analysis showed that the trivial solution 
$X=\mathcal{P},\ Y=0$, where the acoustic subharmonic field is switched-off,
experiences a subcritical pitchfork bifurcation at $\mathcal{P}=1$, the
parametric generation threshold, and gives rise to a finite amplitude
solution $\left\vert X\right\vert =1,$ $\left\vert Y\right\vert ^{2}=(1\pm 
\sqrt{\mathcal{P}^{2}(1+\eta ^{2})-\eta ^{2}})/(1+\eta ^{2}).$
Subcriticality implies a domain of coexistence between the trivial (rest)
state and the finite-amplitude solution when $\mathcal{P}_{T}<$ $\mathcal{P}%
<1,$ where $\mathcal{P}_{T}=\eta /\sqrt{1+\eta ^{2}}$ is the turning point.
For some values of the parameters, the finite amplitude solution can
experience a Hopf (selfpulsing) instability, whose analytical expression was
also given in \cite{Sanchez05}. The boundaries of these local instabilities
are depicted in Fig.1.

\begin{figure}[h]
\centering\includegraphics[width=0.45\textwidth]{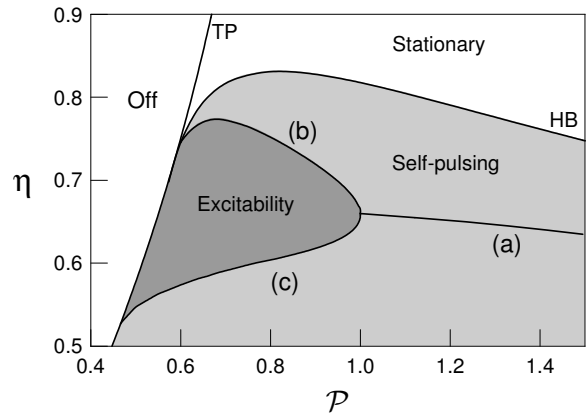}
\caption{Bifurcation diagram of Eqs. (4) for $\protect\gamma =0.1.$ Local
bifurcations are denoted by TP (fold or turning point), HB (Hopf
bifurcation), and the line $\mathcal{P}=1$ corresponding to the subcritical
pitchfork bifurcation (parametric generation threshold). Three curves
labeled as (a), (b) and (c) denote the locus of global bifurcations (see
text for details).}
\end{figure}

In \cite{Sanchez05} it was also shown that the system possess homoclinic
(global) dynamics at a particular value of the nonlinearity parameter $\eta $%
. In this letter we perform a detailed two-parameter numerical analysis of
Eqs. (\ref{eqs2}). For this aim, we reduce the dimensionality of the
parameter space by fixing the value of the relative losses $\gamma =0.1$,
which is in correspondence with typical experimental conditions \cite%
{Brysev00}. The results for other values of $\gamma $ are qualitatively
similar.

The bifurcation diagram shown in Fig. 1 reveals that the system presents a
complex scenario of global dynamics. Besides the Hopf bifurcation leading to
self-pulsing, three different global bifurcations have been identified in
this system. These bifurcations can be detected numerically by computing the
period of the limit cycles, since close to a homoclinic/heteroclinic point
the period diverges to infinity as $T=-\frac{1}{\lambda }\ln (\mathcal{P}%
_{h}-\mathcal{P})$, where $\mathcal{P}_{h}-\mathcal{P}$ measures the
distance to the homoclinic bifurcation (which is assumed small) and $\lambda 
$ is the eigenvalue in the unstable direction of the saddle point \cite%
{Gaspard90}. In Fig. 1, infinite period ($T_{\infty }$) bifurcations are the
curves labeled (a)-(c). Curve (a) corresponds to a gluing (double
homoclinic) bifurcation \cite{Glendinning84}. This bifurcation is
characteristic of systems with $Z_{2}$ symmetry and is mediated by a saddle
point, which in this case corresponds to the trivial state. The gluing
bifurcation exists for a broad range of pump values, and persists until the
pitchfork bifurcation, at $\mathcal{P}=1.$ At this point, two new branches
of $T_{\infty }$ bifurcations emerge. The upper branch (b) correspond to a
homoclinic bifurcation connecting one saddle with itself, while the lower
branch (c) corresponds to a heretoclinic connection between two symmetric
saddles. Three phase portraits corresponding to the different global
bifurcations are shown in Fig. 2. The simultaneous presence of the three
reported types of global bifurcations is a unusual phenomenon. We note that
a similar picture\emph{\ }results from the analysis of two coupled van der
Pol oscillators with delay coupling \cite{Wirkus02}, however in a quite
different context.

\begin{figure}[h]
\centering \includegraphics[width=0.5\textwidth]{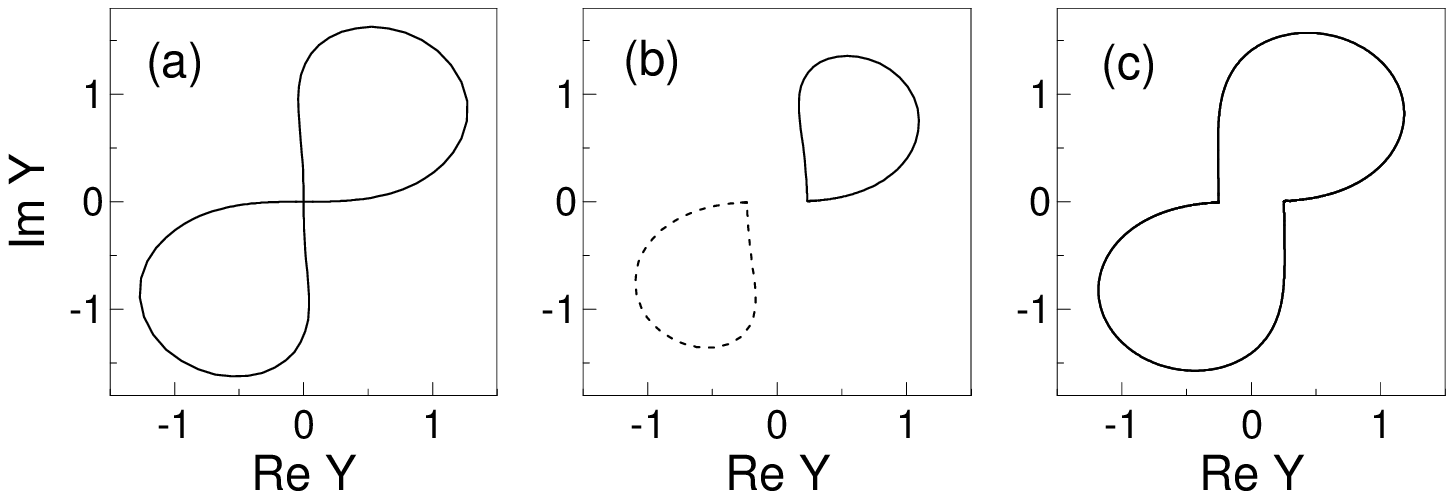}
\caption{Phase portraits of the acoustic field for parameters close to the
different global bifurcations: (a) gluing, for $\protect\eta =0.65$ and $%
\mathcal{P}=1.22$, (b) homoclinic, for $\protect\eta =0.7$ and $\mathcal{P}%
=0.96$, and (c) heteroclinic, for $\protect\eta =0.65$ and $\mathcal{P}=1$. }
\end{figure}

The period of limit cycle solutions, as the pump parameter is varied, is
shown in Fig. 3 for two representative values of the nonlinearity parameter $%
\eta $, below (dashed line) and above (continuous line) the
codimension--three point where all the global bifurcations coalesce, at $%
\eta \approx 0.67$. As expected from Fig. 1, at $\eta =0.7$ there is only
one $T_{\infty }$ bifurcation, consisting in a homoclinic loop as shown in
Fig. 2(b). For $\eta =0.65,$ both gluing bifurcation [Fig. 2(a)] and
heteroclinic connection [Fig. 2(c)] coexist. It is remarkable that this
situation has been also described for a periodically forced Navier-Stokes
flow in hydrodynamics \cite{Lopez00}.

\begin{figure}[h]
\centering\includegraphics[width=0.47\textwidth]{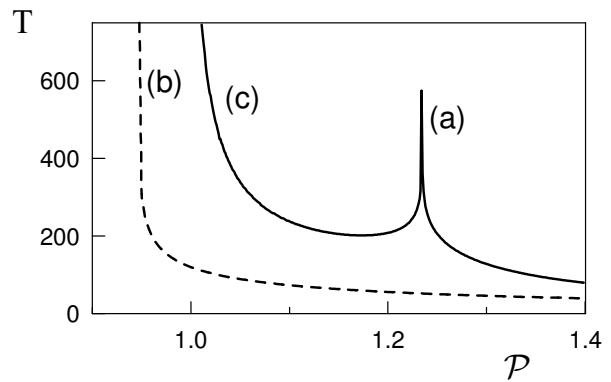}
\caption{Period of time-dependent solutions depending on the pump, for $%
\protect\eta =0.65$ (full line) and $\protect\eta =0.7$ (dashed line). The
period diverges to infinity at some values, denoting the presence of gluing
(a), homoclinic (b) and heteroclinic (c) bifurcations.}
\end{figure}

One prominent property found in some systems presenting homoclinic dynamics
is excitability. As stated above, excitable systems present a highly
nonlinear response to an external stimulus, with a well defined excitability
threshold and a constancy in the reaction when perturbed above threshold. In
addition, a refractory time has to elapse before the system can be excited
again. These neuron--like properties can be also present in the
magnetoacoustic resonator considered here, as we demonstrate below.

The key for achieving excitability is the coexistence of a global
bifurcation and a stable fixed point \cite{Izhikevich00}. This occurs in the
dark-shadowed region in Fig. 1. The excitability properties of a system can
be characterized in several ways, in terms of its respose to different kinds
of external perturbations. In order to demonstrate the main signatures
described above, we consider first the behaviour of the system under a short
(delta-like) perturbation. Figure 4(a) shows the amplitude response of the
acoustic field after four perturbations with increasing amplitude, for
parameter values $\eta =0.7$, $\mathcal{P}=0.94$ and $\gamma =0.1$ (close to
the homoclinic bifurcation boundary, curve (b) in Fig.1). The system,
initially at rest, is perturbed at $t=0.$ The amplitude of the weakest
perturbation was chosen to be below the excitability threshold $y_{th}$
(determined by the distance from the node to the saddle point), and the
system relaxes smoothly to the rest. Three perturbations above the
threshold, with different amplitudes, however generate three identical
pulses or spikes. The amplitude of the perturbation only affects to the
response time of the system: when stronger the stimulus, the smaller the
time needed to develope a pulse.

\begin{figure}[h]
\centering\includegraphics[width=0.42\textwidth]{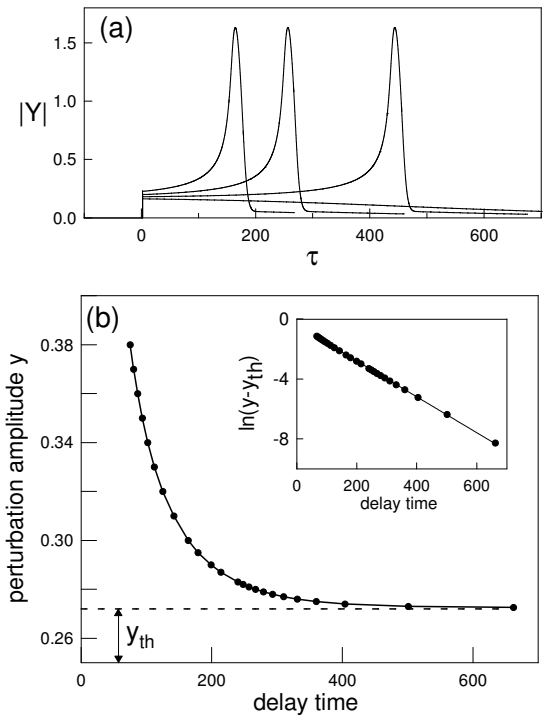}
\caption{Amplitude of the acoustic field in response to four external
perturbations of increasing amplitude, for $\protect\eta =0.7$ and $\mathcal{%
P}=0.94$ (a). Delay time (dots) as a function of the strength of the
perturbation. The inset corresponds to a logaritmic representation of the
numerical data, and the continuous line to Eq. (6). }
\end{figure}

The delay time has been measured as the interval between the stimulus and
the instant where the pulse reaches the maximum amplitude, for different
overthreshold perturbations. The results are shown in Fig. 4(b). The inset
corresponds to the logarithmic representation of the numerical data, and
demonstrates that the scaling law for the response time is ruled by the same
law as the period of limit cycles close to homoclinic/heteroclinic
bifurcations, i.e.%
\begin{equation}
\tau _{delay}=-\frac{1}{\lambda }\ln (y-y_{th})+c,
\end{equation}%
where $\lambda $ again corresponds to the unstable eigenvalue of the saddle
point. For the parameters in Fig. 4, from the linear stability analysis
results $\lambda =0.01165$, in good agreement with the value $\lambda
=0.01176$ obtained from the slope of the linear fit in Fig. 4(b).

Together with the excitation of single spikes, other characteristic features
of many excitable systems are bursting and synchronization phenomena \cite%
{Izhikevich00}. Bursting is the typical firing pattern displayed by neurons,
and consists in the periodic emission of short trains of fast spike
oscillations, intercalated by quiescent intervals. Some systems (e.g. the
neuron) present autonomous bursting, while in some others it can be induced
by a weak periodic modulation of the control parameter, as has been shown in
the CO$_{2}$ laser with feedback \cite{Allaria01}. We solve Eqs. (\ref{eqs2}%
) with a pumping term in the form $\mathcal{P}(t)=\mathcal{P}(1+m\cos
(\omega _{p}t)).$ Different bursting regimes have been observed depending on
the modulation depth $m$ and frequency $\omega _{p}$. Figure 5 shows a
periodic bursting regime for $\mathcal{P}=1,$ $\omega _{p}=0.005$ and $%
m=0.0505$. In this case, one burst is excited every period of the external
driving (1:1 locking). Other modulation parameters result in different $n:m$
phase locking regimes (defined by the ratio of bursting to modulation
periods), or even in chaotic bursting patterns, in agreement with other
externally modulated excitable systems.

\begin{figure}[h]
\centering\includegraphics[width=0.48\textwidth]{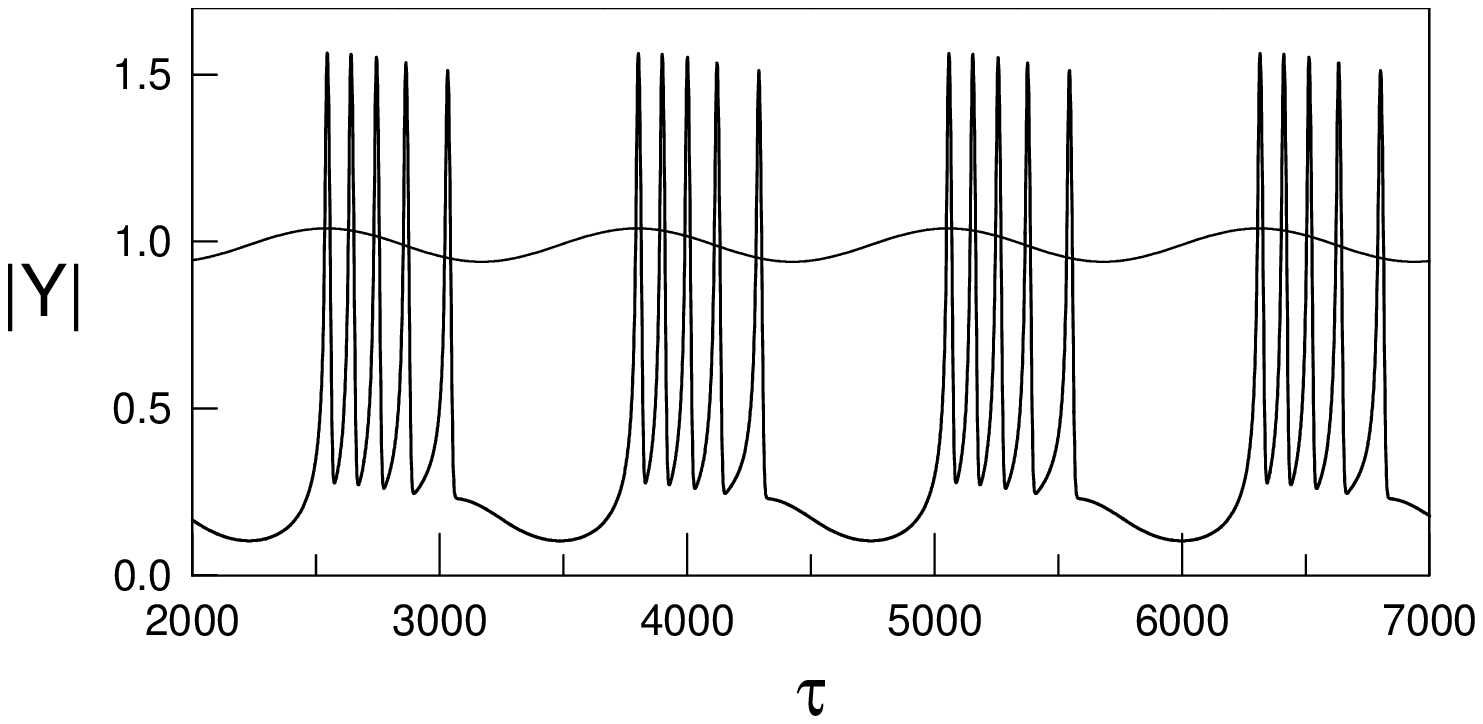}
\caption{Bursting pattern induced by a periodic modulation of the control
parameter. }
\end{figure}

In conclusion, we have presented for the first time an acoustic system
displaying excitability. The system is formed by a magnetostrictive material
excited by an oscillating magnetic field. Some predictions of the
theoretical model reported in this letter are in agreement with recent
experimental results in a nonlinear magnetoacoustic hematite ($\alpha $--Fe$%
_{2}$O$_{3}$) resonator. In particular, bistability \cite{Fetisov06} ans
selfpulsing \cite{Fetisov02}, including oscillating solutions whose period
strongly depends on the pump close to threshold (main signature of
homoclinic dynamics) have been observed.

The work was financially supported by the Spanish Ministerio de Educaci\'{o}%
n y Ciencia, and European Union FEDER (Project FIS2005-07931-C03-03).
Discussions with Y. Fetisov are gratefully acknowledged.

\end{document}